\documentclass{elsart}
\usepackage{graphicx}
\usepackage{epsfig,psfig}
\def\gsim{\;\raise0.3ex\hbox{$>$\kern-0.75em\raise-1.1ex\hbox{$\sim$}}\;}
\def\lsim{\;\raise0.3ex\hbox{$<$\kern-0.75em\raise-1.1ex\hbox{$\sim$}}\;}
\begin{document}

\begin{frontmatter}
\title{{\hfill \small DSF-37-2001, MPI-PhT/2001-51, astro-ph/0111408}\\ $~$\\
A precision calculation of the effective number of cosmological neutrinos}
\author[Napoli]{G.~Mangano},
\author[Napoli]{G.~Miele},
\author[MPI]{S.~Pastor},
\author[Bonn]{M.~Peloso}
\address[Napoli]{Dipartimento di Scienze Fisiche, Universit\`{a} di Napoli
{Federico II} and INFN, Sezione di Napoli, Complesso Universitario di
Monte S. Angelo, Via Cintia, I-80126 Naples, Italy}
\address[MPI]{Max-Planck-Institut f\"{u}r Physik, F\"{o}hringer Ring 6,
D-80805 Munich, Germany}
\address[Bonn]{Physikalisches Institut, Universit{\"{a}}t Bonn,
Nussallee 12, D-53115 Bonn, Germany}

\begin{abstract}
The neutrino energy density of the Universe can be conveniently
parametrized in terms of the so-called effective number of neutrinos,
$N_\nu^{\rm eff}$.  This parameter enters in several cosmological
observables. In particular it is an important input in those numerical
codes, like CMBFAST, which are used to study the Cosmic Microwave
Background anisotropy spectrum. By studying the neutrino decoupling with
Boltzmann equations, one can show that this quantity differs from the
number of massless neutrino species for an additional contribution due to a
partial heating of neutrinos during the $e^{\pm}$ annihilations, leading to non
thermal features in their final distributions. In this paper we review the
different results obtained in the literature and perform a new analysis
which takes into account, in a fully consistent way, the QED corrections at
finite temperature to the photon and $e^{{\pm}}$ plasma equation of state. The
value found for three massless active neutrinos is $N_\nu^{\rm
eff}=3.0395$, in perfect agreement with the recommended value used in
CMBFAST, $N_\nu^{\rm eff}=3.04$. We also discuss the case of additional
relativistic relics and massive active neutrinos.
\end{abstract}
\end{frontmatter}

\section{Introduction}

\label{sec:introduction}

At temperatures below the muon mass and above $\sim 10$ MeV, the
Universe is filled by a plasma of photons, electrons, positrons, and
neutrinos, kept in thermodynamical equilibrium by the electroweak
interactions. As the temperature drops below this value, the rate of
weak interactions starts to be comparable with the Universe expansion
rate, and neutrinos decouple from the electromagnetic
$\gamma,\,e^{\pm}$ plasma. For most practical purposes, it is accurate
enough to consider the freeze-out of neutrinos as fully achieved at
temperatures of about $2-3$ MeV. In this limit neutrinos do not share
any entropy release from $e^{\pm}$ annihilations, once the temperature
drops further, below the electron mass. Assuming that all the entropy
produced by the annihilations is transferred to photons, their
temperature $T$ is increased with respect to the neutrino temperature
by the well known factor $T/T_\nu = \left( 11/4 \right)^{1/3}$ (see,
for example, \cite{kt}). Actually more accurate calculations show that
neutrinos are still slightly interacting with the electromagnetic
plasma and thus receive a small portion of the entropy from $e^{\pm}$
annihilations~\cite{dicus}. Neutrinos with higher momenta are more
heated, since, in the relevant range of energies, weak interactions
get stronger with rising energy. This produces a momentum dependent
distortion in the neutrino spectra from the equilibrium Fermi--Dirac
behaviour.

A further, though smaller, effect on $T/T_\nu$ is induced by finite
temperature QED corrections to the electromagnetic plasma. In fact
electromagnetic interactions modify the $e^{\pm}$ and $\gamma$ dispersion
relations, and thus the energy density and pressure of the plasma. More
precisely, the energy density is lowered so that the $e^{\pm}$ annihilation
phase releases less entropy with respect to the non-interacting particle
limit calculation. Since most of this energy ends up into photons, this
decrease results in a smaller $T/T_\nu$ ratio \cite{heckler}.

While any direct observation of the actual distortions in the neutrino
distributions is out of question, nevertheless we can hope that the
increase of the total energy of the relic neutrinos may have a sizeable
effect on the expansion rate of the Universe. The energy density stored in
relativistic species, $\rho_R$, is customarily given in terms of the
so-called {\it effective number of neutrino species} \cite{Sh69,St77},
$N_\nu^{\rm eff}$, through the relation
\begin{equation}
\rho_{\rm R} = \left[ 1 + \frac{7}{8} \left( \frac{4}{11}
\right)^{4/3} \, N_\nu^{\rm eff} \right] \, \rho_\gamma \,\,,
\label{neff}
\end{equation}
where $\rho_\gamma$ is the energy density of photons, whose value today is
known from the measurement of the Cosmic Microwave Background (CMB)
temperature. Eq. (\ref{neff}) can be also written as
\begin{equation}
N_\nu^{\rm eff} \equiv \left( \frac{\rho_R -\rho_\gamma}{\rho^0_\nu}\right)
\left(\frac{\rho^0_\gamma}{\rho_\gamma} \right)\, ,
\label{neff-def}
\end{equation}
where $\rho_\nu^0$ denotes the energy density of a single specie of
massless neutrino with an equilibrium Fermi-Dirac distribution, and
$\rho^0_\gamma$ is the photon energy density in the instantaneous neutrino
decoupling approximation. The normalization of $N_\nu^{\rm eff}$ is such
that it gives $N_\nu^{\rm eff}
= 3$ in the standard case of three families of massless neutrinos,
in the limit of an instantaneous decoupling. In principle, $N_\nu^{\rm
eff}$ can receive contribution from other relativistic relics with energy
density $\rho_X$ as well. In the following we will mostly restrict our
analysis to the standard case, but we will further consider this more
general framework in Section 4.

As we will discuss below, when considered separately, the non instantaneous
neutrino decoupling gives $\Delta N_\nu^{\rm eff} \equiv N_\nu^{\rm eff} -
3 \simeq 0.034\,$~\cite{dhs,dhs2,emp}, while QED effects contribute for
about $\Delta N_\nu^{\rm eff} \simeq 0.011\,$~\cite{heckler}. Could the two
effects be added linearly, they therefore would produce a final value
$N_\nu^{\rm eff}
\simeq 3.045$. Recently, these effects have been reconsidered
in~\cite{steigman}. This work combines the results for the non
instantaneous neutrino decoupling obtained by a numerical calculation in
\cite{gnedin} and then replacing the ratio $T/T_\nu=\left( 11/4
\right)^{1/3}$ with the value obtained by considering QED corrections.
This procedure may provide a good estimate, but it is worth pointing out
that, since QED effects modify several points of the Boltzmann equations
describing neutrino decoupling, it is advisable to study any possible
interplay of the two effects. Actually a precision calculation of this
kind, where both these effects are included at the same time, is still
lacking, and it is the aim of the present paper. We will show that this
interplay ends up in a $10 \%$ smaller total correction to $N_\nu^{\rm
eff}$ than what would be obtained by simply adding the two contributions.
We also notice that the calculation of~\cite{gnedin} seems to
underestimate\footnote{In ref.~\cite{gnedin} it is claimed that their
result is enhanced with respect to the one of~\cite{dhs}. We think that
this is actually due to a misinterpretation of the results of~\cite{dhs} in
terms of $N_\nu^{\rm eff}\,$. This can be easily understood by noticing
that in ~\cite{gnedin} it is found a final ratio $T/T_\nu$ closer to
$\left( 11/4\right)^{1/3}$ than what obtained in ~\cite{dhs,dhs2,emp}.} the
corrections from non instantaneous neutrino decoupling with respect to the
result found in \cite{dhs,dhs2,emp}. Our analysis is based on a numerical
code described in \cite{emp}, which has been modified to take into account
QED finite temperature effects.

{}From the observational point of view the effects considered in this paper
are too small to influence Big Bang Nucleosynthesis (BBN), since they
produce a change in the $^4$He mass fraction, $\Delta Y(^{4}\mbox{He}) \sim
10^{-4}$ ~\cite{emp,fdt,lotu}, which is smaller than the actual
theoretical, $5~10^{-4}$ \cite{lotu,Esposito:2000hh}, and experimental,
$\gsim 2~10^{-3}$ \cite{izotov}, uncertainties on this quantity. This
change is actually slightly larger when one takes into account the effect
of flavour neutrino oscillations, as recently shown in
\cite{Hannestad:2001iy}.  More promisingly, they might be detected via
future precision CMB anisotropy measurements at high multipoles, since
cosmic variance prevents their resolution on scales probed by present
balloon experiments. According to a recent analysis \cite{cmb} (see also
\cite{Esposito:2000hh,Hannestad:2000hc,Orito:2000zb,Esposito:2000sv,Kneller:2001cd,Hannestad:2001hn}),
CMB temperature measurements by the Planck satellite experiment will be
able to provide a measure of the relativistic energy density with a
precision of about $\Delta N_\nu^{\rm eff} \simeq 0.2\,$, without any
strong prior on the other cosmic parameters.  Furthermore, as discussed
in~\cite{cmb2}, the situation is more promising if polarization
measurements will be available, and some stronger priors are imposed on
other cosmological parameters, enforced for example by independent
measurements. In this case, $\Delta N_\nu^{\rm eff}$ would be determined
with an accuracy comparable or, according to~\cite{cmb2}, even higher than
the order of magnitude of the effects we are here considering.  We notice
that in the CMB data analysis the presence of a non vanishing $\Delta
N_\nu^{\rm eff}$ is already considered with, for example, a recommended
default value for three massless active neutrinos $N_\nu^{\rm eff}=3.04$ in
the CMBFAST code \cite{SZ} . A precise check of this important parameter is
therefore mandatory.

The paper is organized as follows. In Section \ref{sec:nuheating} we
summarize the set of equations and our numerical approach to compute the
neutrino distortion due to their incomplete decoupling during the $e^{\pm}$
annihilation phase. In Section \ref{sec:qed} we also consider the
corrections to the equation of state of $e^{{\pm}}$ and $\gamma$ plasma due to
QED effects at finite temperature, showing how these effects modify the
numerical computation of Section \ref{sec:nuheating}, and we present our
results. We also discuss the general case of extra relativistic degrees of
freedom and of massive active neutrinos in Section \ref{sec:extra}. Finally
our conclusions are reported in Section \ref{sec:concl}.

\section{Corrections from to the non instantaneous decoupling}
\label{sec:nuheating}

The effects of non-instantaneous neutrino decoupling have been addressed in
several studies, either based on analytical methods~\cite{dicus}, on
numerical analysis with some simplifying approximations
\cite{herrera,rana,dt,df}, or finally on full numerical computations, which
solve the Boltzmann equations until neutrinos are completely decoupled
\cite{dhs,dhs2,emp,gnedin,madsen}.

The procedure is easily summarized. Following the notation
of~\cite{dhs,emp}, a convenient choice for the {\it time} variable is
given by $x \equiv m \, R$, where $R(t)$ is the scale factor of the
Universe (chosen to have dimension of length) and $m$ some reference
mass, taken to be the electron mass as in~\cite{emp}. We also introduce
the dimensionless comoving momentum $y \equiv p \, R\,$, and the
rescaled photon temperature $z \equiv T \, R\,$. In this notation, the
Boltzmann equations for the neutrino distributions can be written in
the form
\begin{equation}
\frac{d}{d x} \, f_{\nu_\alpha} \left( x,y \right) = \frac{1}{x\,H} \,
I_{\nu_\alpha} \, \left[ f_{\nu_e}, f_{\nu_x} \right] \,\,,
\label{boltz}
\end{equation}
where $H$ is the Hubble parameter, while $I_{\nu_\alpha}$ represents the
collisional integral, in momentum space $y$, for the single neutrino specie
$\nu_\alpha$, and is a functional of all neutrinos and electron/positron
distributions\footnote{Since $e^{\pm}$ are kept in thermodynamical equilibrium
with $\gamma$ by the electromagnetic interactions, they have Fermi--Dirac
distributions and share the same temperature of photons, $T$. We neglect
the completely irrelevant $e^{\pm}$ asymmetry.}. For the processes of interest
(see~\cite{dhs} for the detailed calculation), some of the integrations
appearing in the $I_{\nu_\alpha}$ can be analytically performed, and the
collisional integrals can be reduced to two-dimensional integrals. In the
range of temperatures we are interested in, electron neutrinos experience
charged current interactions due to the presence of $e^{\pm}$ in the thermal
bath, while all active neutrino flavours interact via neutral current
interactions. For this reason, the nonequilibrium corrections to the
distribution function $f_{\nu_e}$ are different from those of the other two
neutrino species $f_{\nu_x}$ ($x$ denoting both $\mu$ and $\tau$).

The two Boltzmann equations for $f_{\nu_e}$ and $f_{\nu_x}$ must be
supplemented by the continuity equation
\begin{equation}
\frac{d}{d x} {\bar \rho} \left( x \right) = \frac{1}{x}
\left( {\bar \rho} - 3 \, {\bar P} \right) \,\,,
\label{energy}
\end{equation}
where ${\bar \rho} = \rho \left( x/m_e \right)^4 \,$, ${\bar P} = P \left(
x/m_e \right)^4 \,$, and $\rho$ and $P$ are the energy density and pressure
of the $\gamma,\,e^{\pm},\,\nu$ plasma.  Eq.~(\ref{energy}) can be rewritten as
an evolution equation for the quantity $z\,$ which gives the ratio
$T/T_\nu$. In case of instantaneous neutrino decoupling, the asymptotic
value of $z$, denoted by $z^{0}\,$, results to be the well-known value
$(11/4)^{1/3}$.

As in Ref.~\cite{emp} the unknown neutrino distributions are parametrized as
\begin{equation}
f_{\nu_\alpha} \left( x,y \right) = \frac{1}{{\rm e}^y +1} \, \left[ 1
+ \sum_{i=0}^\infty\,
a_{i}^\alpha(x)\, P_i \left( y \right) \right]  \,,
\label{expan}
\end{equation}
where $P_i(y)$ are orthonormal polynomials with respect to the Fermi
function weight
\begin{equation}
\int_0^\infty \frac{dy}{{\rm e}^{y}+1} \, P_i \left( y \right) \, P_j
\left( y \right) = \delta_{ij} \,\,.
\label{ortho}
\end{equation}
By substituting Eq.~(\ref{expan}) into Eqs. (\ref{boltz}), one gets
\begin{equation}
\frac{d}{d x}\, a_{i}^\alpha(x) = \frac{1}{x H}\, \int_0^\infty
dy_1\, P_i \left( y_1 \right) \, I_{\nu_\alpha}\left[
f_{\nu_e},f_{\nu_x} \right] \,\,.
\label{eqc}
\end{equation}
Since at high temperature the neutrinos are in thermal equilibrium, the
initial condition for the coefficients is $a_{i}^\alpha=0$.

To perform the numerical computation it is necessary to truncate the
infinite series in Eq.~(\ref{expan}) at some finite value $n$, where the
choice of $n$ depends on the accuracy we do require on the results; it is
found in~\cite{emp} that $n=3$ gives an accuracy of about $1 \%$ in the
neutrino distortions. In this case, the asymptotic value found for $z$,
denoted by $z^{\rm fin}$, slightly differs from the instantaneous
decoupling result $z^{0}$
\begin{equation}
\frac{\delta z}{z^{0}} \equiv \frac{z^{\rm fin} - z^{0}}{z^{0}}= - 1.406
{\cdot} 10^{-3}\, \,.
\label{deltaz}
\end{equation}
The final neutrino distribution functions show a non thermal behaviour due
to the presence of non vanishing ripple terms. Having fixed $n=3$, we
rewrite the expression of Eq. (\ref{expan}) in the form
\begin{equation}
f_{\nu_\alpha}^{\rm fin} \left(y \right) = \frac{1}{{\rm e}^y +1} \,
\left( 1 +  10^{-4} \sum_{i=0}^{3} c^\alpha_i~y^i
\right)\, \,,
\label{simple-expan}
\end{equation}
and report the final values for the coefficients $c^\alpha_i$ in Table
\ref{coeff}. By using Eq. (\ref{simple-expan}) it is now immediate to
compute the additional contribution to the neutrino energy density due to
incomplete decoupling. Defining
\begin{table*}
\begin{center}
\begin{tabular}{ccccc}
\hline
Flavour ($\alpha$)& $c^\alpha_0$ & $c^\alpha_1$ & $c^\alpha_2$ &
$c^\alpha_3$\\ \hline $e$& -2.556& -2.739& 6.133& -0.1477\\ $\mu,\tau$&
-2.049& -2.259& 3.145& -0.1129\\ \hline
\end{tabular}
\end{center}
\caption{Values of the coefficients of
Eq.~(\ref{simple-expan}) from \cite{emp} (QED effects not included).}
\label{coeff}
\end{table*}
\begin{eqnarray}
\frac{\delta \rho_{\nu_\alpha}}{\rho_\nu^0} & \equiv &
\frac{\int_{0}^{\infty} dy~y^3~f_{\nu_\alpha}^{\rm fin}(y)}{\int_{0}^{\infty}
dy~y^3~\left({\rm e}^y +1
\right)^{-1}} - 1 \nonumber\\ &=& 10^{-4}\left( c^{\alpha}_0 + 2700
\left(\frac{\zeta(5)}{7 \pi^4}\right) c^{\alpha}_1 + 310 \left(
\frac{\pi^2}{147} \right) c^{\alpha}_2 + 12150 \left(
\frac{\zeta(7)}{\pi^4} \right) c^{\alpha}_3 \right) \, \, , \nonumber \\
\label{deltarho}
\end{eqnarray}
one obtains, for the values of Table \ref{coeff}, $\delta \rho_{\nu_e} /
\rho_{\nu}^0 = 0.953 \%$ and $\delta \rho_{\nu_x} / \rho_{\nu}^0 =
0.399\%$. Finally, from the definition of $ N_\nu^{\rm eff}$ of Eq.
(\ref{neff}), it is straightforward to get the following expression
\begin{equation}
N_\nu^{\rm eff} = \left( \frac{z^{0}}{z^{\rm fin}}\right)^4 \left( 3
+ \frac{\delta \rho_{\nu_e}}{\rho_\nu^0} + 2 \frac{\delta
\rho_{\nu_x}}{\rho_\nu^0} \right) \simeq \left( 3-12 \frac{\delta z}{z^0} +
\frac{\delta \rho_{\nu_e}}{\rho_\nu^0} + 2 \frac{\delta
\rho_{\nu_x}}{\rho_\nu^0} \right)\,\, ,
\label{neff-emp}
\end{equation}
which gives the {\it effective number} for three massless active neutrinos
for temperatures below the neutrino decoupling ($\sim 1$ MeV). The values
found from the numerical analysis then give $N_\nu^{\rm eff} = 3.0345$,
which fixes the additional contribution to be $\Delta N_\nu^{\rm eff} =
0.0345$. Notice that from Eq. (\ref{neff-emp}),   $\Delta N_\nu^{\rm eff}$
takes two contributions. The term proportional to $\delta z/z^{0}$ accounts
for the smaller profit that photon temperature gets from $e^{\pm}$
annihilations, whereas the contributions due to $\delta
\rho_{\nu_{e,x}}/\rho^0_\nu$ are a measure of the non thermal behaviour
of neutrino distribution functions. These two terms are indeed of the same
size.

Remarkably, our results are in very good agreement with a previous analysis
performed applying a different numerical technique \cite{dhs,dhs2}. They
find $\delta z/z^{0} = (-1.37 {\pm} 0.02) {\cdot} 10^{-3}$, and an  increase of
energy with respect to the instantaneous decoupling case of $\delta
\rho_{\nu_e} / \rho_\nu^0 = \left( 0.946 {\pm} 0.001 \right) \%$ for the
electron neutrinos and of $\delta \rho_{\nu_x} /
\rho_{\nu}^0 = \left( 0.398 {\pm} 0.001 \right) \%$ for the other two
species. These values give the effective number of neutrino species
$N_\nu^{\rm eff} = 3.0340 {\pm} 0.0003\,$.

On the other hand we disagree with the results of \cite{gnedin}, where by
solving the same Boltzmann equations it is found $\delta
\rho_{\nu_e} /
\rho_{\nu}^0 = 0.607 \%$,  $\delta \rho_{\nu_x} / \rho_{\nu}^0 = 0.256 \%$
and  $\delta z/z^{0}=-0.888 {\cdot} 10^{-3}\,$, which finally gives $N_\nu^{\rm
eff} = 3.022$. In both calculations~\cite{dhs,dhs2} and~\cite{gnedin}, the
integral--differential Boltzmann equations are solved through a grid in
momentum space. The grid adopted in~\cite{gnedin} extends to a larger
momentum range than the one of~\cite{dhs,dhs2}. The grid used
in~\cite{dhs,dhs2} is however denser in the interval $0.1 < y \,
\left( m/ {\rm MeV} \right) < 20\,$. For neutrino distributions in thermal
equilibrium, the $97.5 \%$ of the energy density comes from particles with
momentum in the range $1 <  y \, \left( m/ {\rm MeV} \right) <10\,$.
According to the analysis of~\cite{dhs2}, the different results
of~\cite{gnedin} may be due to a lack of precision in this most relevant
interval. To conclude, we can only point out that the beautiful agreement
of our findings with what obtained in \cite{dhs,dhs2}, despite of the
rather different numerical method, make us rather confident on our result
for $N_\nu^{\rm eff}$.

\section{Corrections from QED at finite temperature}
\label{sec:qed}

Finite temperature QED corrections modify in several points the
calculations of neutrino decoupling described till now. First, through the
change on the electromagnetic plasma equation of state, they affect Eq.
(\ref{energy}). Moreover, since $e^{\pm}$ masses are renormalized by a finite
temperature term, the r.h.s. of Eq.s (\ref{boltz}) and (\ref{eqc}) should
be modified correspondingly. Finally, the change of energy density modifies
the expansion rate $H$ in the Boltzmann equations (\ref{boltz}). The
effects of these corrections on neutrino decoupling temperature have been
considered in \cite{nicolao}. Their analysis is however in the framework of
the instantaneous neutrino decoupling limit, therefore they do not consider
the non thermal effects described in the previous Section.

The change in the electromagnetic plasma equation of state can be evaluated
by first considering the corrections induced on the $e^{\pm}$ and photon
masses. They can be obtained perturbatively by computing the loop
corrections to the self-energy of these particles.  For the
electron/positron mass, up to order $\alpha \equiv e^2/\left( 4 \, \pi
\right)$ we find the additional finite temperature
contribution~\cite{heckler}~\footnote{Our result agrees with Eq. (B2)
reported in~\cite{heckler}, while we think that the analogous result in Eq.
(35) of \cite{lotu} contains some misprints.}
\begin{eqnarray}
\delta m_e^2 \!\!\!&&\!\!\! \left( p, \,T \right) =
\frac{2\,\pi\,\alpha\,T^2}{3} + \frac{4\,\alpha}{\pi} \, \int_0^\infty
d k \, \frac{k^2}{E_k} \, \frac{1}{{\rm e}^{E_{k}/T}+1} \nonumber\\
&-& \frac{2 \, m_e^2 \,\alpha}{\pi \, p} \, \int_0^\infty d k \,
\frac{k}{E_{k}} \, \log\left| \frac{p+k}{p-k} \right| \frac{1}{{\rm
e}^{E_{k}/T}+1} \,\,,
\label{dme}
\end{eqnarray}
where $E_{k} \equiv \sqrt{k^2 + m_e^2}$. While the first two terms of this
expression depend on the plasma temperature $T$ only, the last one depends
on the $e^{\pm}$ momentum $p\,$ as well. However, by averaging this term over
the equilibrium $e^{\pm}$ distribution, one easily finds that it contributes
for less than $10\,\%$ to $\delta m_e^2\,$.  For this reason we will
neglect it in the following (see also~\cite{lotu}).

The renormalized photon mass in the electromagnetic plasma is instead
given, up to order $\alpha$, by~\cite{nicolao}
\begin{equation}
\delta m_\gamma^2 \left( T \right) = \frac{8\,\alpha}{\pi} \,
\int_0^\infty d k \, \frac{k^2}{E_k} \, \frac{1}{{\rm e}^{E_{k}/T}+1}
\,\,.
\label{dmg}
\end{equation}
The corrections~(\ref{dme}) and (\ref{dmg}) modify the corresponding
dispersion relations as $E_i^2 = k^2 + m_i^2 + \delta m_i^2 \left( T
\right)\,$ ($i = e,\, \gamma$). This in turn affects the total
pressure and the energy density of the electromagnetic plasma
\begin{eqnarray}
P &=& \frac{T}{\pi^2} \, \int_0^\infty d k \, k^2 \, \,
\log\left[\frac{\left( 1 + {\rm e}^{-E_e/T} \right)^2}{ \left( 1 -
{\rm e}^{-E_\gamma/T} \right)} \right]\,\,,\\ \rho &=& - P + T \,
\frac{d P}{d T} \,\,.
\label{energ}
\end{eqnarray}
Expanding $P$ with respect to $\delta m_e^2$ and $\delta m_\gamma^2$, one
obtains the first order correction
\begin{eqnarray}
P^{\rm int} &=& - \int_0^\infty \frac{dk}{2 \, \pi^2} \left[
\frac{k^2}{E_k} \, \frac{\delta m_e^2 \left( T \right)}{{\rm
e}^{E_{k}/T}+1} + \frac{k}{2} \, \frac{\delta m_\gamma^2 \left( T
\right)}{{\rm e}^{k/T}-1} \right] \,\,.
\label{pint}
\end{eqnarray}
Note that an important factor $1/2$ must be introduced for not double
counting the renormalization effect on the total pressure, which now reads
$P=P^0+ P^{\rm int}$ \cite{heckler}, $P^0$ being the value of the pressure
for the non-interacting particle gas. The energy density is then obtained
by using $P$ in Eq. (\ref{energ}).

\begin{table*}
\begin{center}
\begin{tabular}{ccccc}
\hline
Flavour ($\alpha$)& $c^\alpha_0$ & $c^\alpha_1$ & $c^\alpha_2$ &
$c^\alpha_3$\\ \hline $e$& -2.507& -2.731& 6.010& -0.1419\\ $\mu,\tau$&
-2.003& -2.196& 3.061& -0.1091\\ \hline
\end{tabular}
\end{center}
\caption{Values of the coefficients of
Eq.~(\ref{simple-expan}) when QED effects are included.}
\label{coeff2}
\end{table*}

The presence of the additional contributions $P^{\rm int}$ and $\rho^{\rm
int}$ modify the evolution equation for $z$ contained in Eq.
(\ref{energy}), which now reads
\begin{equation}
\frac{dz}{dx} = \frac{\frac{x}{z}J(x/z)-
\frac{1}{2 \pi^2 z^3}\int_0^\infty~dyy^3\left(\frac{df_{\nu_e}}{dx}+
2\frac{df_{\nu_x}}{dx}\right)+G_1(x/z)}
{\frac{x^2}{z^2}J(x/z)+Y(x/z)+\frac{2\pi^2}{15}+G_2(x/z)}\, ,
\label{dzdx}
\end{equation}
where
\begin{eqnarray}
G_1(\omega) &=& 2 \pi \alpha \left[ \frac{1}{\omega}
\left(\frac{K(\omega)}{3} + 2 K(\omega)^2-
\frac{J(\omega)}{6}- K(\omega)~J(\omega) \right) \right. \nonumber \\
&+& \left.  \left(
\frac{K'(\omega)}{6}-K(\omega)~K'(\omega)+\frac{J'(\omega)}{6} +
J'(\omega)~K(\omega) + J(\omega)~K'(\omega)\right ) \right] \nonumber \\ &&
\\ G_2(\omega) &=&
-8
\pi
\alpha
\left( \frac{K(\omega)}{6} + \frac{J(\omega)}{6} - \frac{1}{2} K(\omega)^2 +
K(\omega)~J(\omega)\right)
\nonumber\\
+2 \pi \alpha &\omega & \left(\frac{K'(\omega)}{6}-K(\omega)~K'(\omega)
+\frac{J'(\omega)}{6} + J'(\omega)~K(\omega) + J(\omega)~K'(\omega)\right)
\nonumber \, \, , \\
&&
\end{eqnarray}
with
\begin{eqnarray}
K(\omega) & = & \frac{1}{\pi^2}~\int_0^\infty
du~\frac{u^2}{\sqrt{u^2+\omega^2}}~
\frac{1}{\exp\left(\sqrt{u^2+\omega^2}\right)+1} \, , \\
J(\omega) & = &  \frac{1}{\pi^2}~\int_0^\infty
du~u^2~\frac{\exp\left(\sqrt{u^2+\omega^2}\right)}
{\left(\exp\left(\sqrt{u^2+\omega^2}\right)+1\right)^2} \, , \\
Y(\omega) & = &  \frac{1}{\pi^2}~\int_0^\infty
du~u^4~\frac{\exp\left(\sqrt{u^2+\omega^2}\right)}
{\left(\exp\left(\sqrt{u^2+\omega^2}\right)+1\right)^2} \, .
\end{eqnarray}
The functions $K'(\omega)$ and $J'(\omega)$ stand for the first derivative
of $K(\omega)$ and $J(\omega)$ with respect to their argument. Note that
neutrinos affect the final value of $z$ through the terms $df_{\nu_x}/dx$,
which are not vanishing only if neutrinos have a non thermal behaviour.

In case one neglects the finite temperature QED corrections, the functions
$G_1(x/z)$ and $G_2(x/z)$ in (\ref{dzdx}) vanish, and one recovers the
expression reported in \cite{emp}. Notice that the presence of $G_2(x/z)$
in the denominator of the r.h.s. of (\ref{dzdx}) makes, at least in
principle, not correct to simply sum the neutrino contribution with the QED
one.

Once both effects, neutrino incomplete decoupling and QED corrections to
electromagnetic plasma, are included into the code of Ref.\cite{emp}, we
find the new results $\delta \rho_{\nu_e} / \rho_{\nu}^0
= 0.935 \%$, $\delta \rho_{\nu_x} / \rho_{\nu}^0 = 0.390 \%$ and $\delta
z/z^{0}=-1.841 {\cdot} 10^{-3}$ which give $N_\nu^{\rm eff} = 3.0395$. The values
of the $c^\alpha_i$ coefficients can be found in Table 2. In Table 3 we
report a comprehensive summary of our results. Comparing  the last two
columns of this table, we see that introducing the finite temperature QED
corrections in the non instantaneous decoupling scenario, leads to a change
on the effective number of neutrinos of $0.005\,$, which is a factor $2$
less than what has been obtained in Refs. \cite{heckler,lotu}. This is
actually due to the interplay of incomplete decoupling of neutrinos and
plasma effect (see the r.h.s of Eq. (\ref{dzdx})), which therefore, at the
level of accuracy we are considering, cannot be considered separately and
then added linearly. Notice that the increase of $N_\nu^{\rm eff}$ is
mainly due to a variation of ${\delta z}/z^{0}$, the changes on $\delta
\rho_{\nu_\alpha}/\rho_\nu^0$ being much smaller. Remarkably, the overall result
$N_\nu^{\rm eff} = 3.0395$ turns out to be in excellent agreement with the
recommended default value $N_\nu^{\rm eff}=3.04$ used in the CMBFAST code
\cite{SZ}.

\begin{table*}
\begin{center}
\begin{tabular}{ccccc}
\hline
  & \cite{dhs,dhs2} (no-QED) & \cite{gnedin} (no-QED) & \cite{emp} (no-QED) &
our work (QED)\\ \hline
$\delta z/z^{0}$           & $(-1.37 {\pm} 0.02) 10^{-3}$ & $-0.888 {\cdot} 10^{-3}$ &
 $-1.406 {\cdot} 10^{-3}$ &  $-1.841 {\cdot} 10^{-3}$ \\
$\delta \rho_{\nu_e}/\rho_{\nu}^0$ & $(0.946 {\pm} 0.001)\%$ & $0.607\%$ &
$0.953\%$ &$0.935\%$ \\ $\delta \rho_{\nu_x}/\rho_{\nu}^0$ & $(0.398 {\pm}
0.001)\%$ & $0.256\%$ & $0.399\%$ & $0.390\%$ \\ $N_\nu^{\rm eff}$
& $3.0340 {\pm} 0.0003 $&$3.022$ &$3.0345$ & $3.0395$\\
\hline
\end{tabular}
\end{center}
\caption{The results of the different analyses.}
\label{summary}
\end{table*}

\section{Extra relativistic degrees of freedom and massive neutrinos}
\label{sec:extra}

We now consider the possibility that, at the stage of neutrino decoupling,
there are extra relativistic degrees of freedom, provided by some $X$ field
excitations, which are assumed to have a thermal distribution with some
temperature $T_X$. Their contribution $\rho_X$ to the total energy density
of the Universe can be parametrized in terms of an additional contribution
$\Delta N_\nu^{\rm eff}$ in the effective number of neutrinos, as defined
in eq. (\ref{neff-def})
\begin{equation}
\Delta N_\nu^{\rm eff} = \left( \frac{z^0}{z^{\rm fin}} \right)^4 N_X \, \, ,
\label{deltan}
\end{equation}
where
\begin{equation}
N_X= \frac{4}{7} g_X \left( \frac{11}{4} \right)^{4/3} \left(
\frac{z_X}{z^0}
\right)^{4} \, \, ,
\label{nx}
\end{equation}
and we have defined $z_X= T_X R$. The parameter $g_X$ depends on the spin
($g_X=1$ for a real scalar, $g_X=7/4$ for a Weyl spinor, etc.) as well as
on the additional internal degrees of freedom of the $X$ particles. Notice
that if the $X$ excitations have decoupled  between $\mu^{\pm}$ and $e^{\pm}$
annihilation phases we simply have $N_X= 4/7 g_X$. For an earlier
decoupling we have instead $N_X < 4/7 g_X$.

The presence of $\rho_X$, apart from introducing a new contribution to
$N_\nu^{\rm eff}$, slightly affects the results obtained in the previous
Sections, namely the relative change of neutrino energy density $\delta
\rho_{\nu_{e,x}} / \rho^0_{\nu}$ induced by incomplete neutrino decoupling,
as well as the asymptotic photon temperature $z^{\rm fin}$. In fact since
their energy density increases the expansion rate of the Universe, we
expect $\delta \rho_{\nu_{e,x}}
/ \rho^0_{\nu}$ to decrease with growing $N_X$, and the ratio
$z^0/z^{\rm fin}$ to become closer to unity, since neutrino decoupling
process starts at earlier times. If we denote with $\delta \rho_{\nu_{e,x}}
(N_X)/
\rho^0_{\nu}$ and $z^{\rm fin}(N_X)$ the new values for these parameters as
functions of $N_X$ we therefore have
\begin{equation}
N_\nu^{\rm eff} = \left( \frac{z^{0}}{z^{\rm fin}(N_X)}\right)^4 \left( 3 +
\frac{\delta \rho_{\nu_e}(N_X)}{\rho_\nu^0} + 2 \frac{\delta
\rho_{\nu_x}(N_X)}{\rho_\nu^0} + N_X\right) \, \, .
\label{neffx}
\end{equation}
Since we are interested to those contributions to $N_\nu^{\rm eff}$
corresponding to species which are relativistic for temperatures in the
$MeV$ range, we can severely bound $\Delta N_\nu^{\rm eff}$ using the
results from BBN, which leads to the conservative bound $\Delta N_\nu^{\rm
eff} \leq 1$ (see for example \cite{bbn} for a recent analysis). Using our
numerical code we have evaluated how $z^{\rm fin}(N_X)/z^0$ and $\delta
\rho_{\nu_e}(N_X)/ \rho_\nu^0$ change when $N_X$ varies in the range $0<N_X<1$.
In this interval, with an accuracy of $10^{-4}$ on $N_\nu^{\rm eff}$, we
find
\begin{eqnarray}
N_\nu^{\rm eff} &=& \left( \frac{z^{0}}{z^{\rm fin}}\right)^4 \left( 3 +
\frac{\delta \rho_{\nu_e}}{\rho_\nu^0} + 2 \frac{\delta
\rho_{\nu_x}}{\rho_\nu^0} +  (1-\beta)  N_X \right) \nonumber \\
&=& 3.0395 + (1-\beta)
\left( \frac{z^{0}}{z^{\rm fin}}\right)^4 N_X \, \, ,
\label{neffxapprox}
\end{eqnarray}
with $\beta= 0.0014$ and where now both $z^0/z^{\rm fin}$ and $\delta
\rho_{\nu_{e,x}}/\rho_\nu^0$ in this expression are those reported in the last
column of Table 3, i.e. with $N_X=0$. The changes of these parameters with
$N_X$ is encoded in an additional term, which is weighted by the small
parameter $\beta$.

It should be clear from what we said before that only species which are
relativistic at the neutrino decoupling, down to the $e^{{\pm}}$ annihilation
stage, contributes for this extra term. It is known that the effective
number of neutrinos at later stages, as for example at recombination, is
much less constrained from data
\cite{cmb,Hannestad:2000hc,Orito:2000zb,Esposito:2000sv,Kneller:2001cd,Hannestad:2001hn,bbn,cmb3}
and it is still possible the case that energy density in the form of
relativistic plasma may be injected only well after the Big Bang
Nucleosynthesis epoch. This implies that the value of $N_\nu^{\rm eff}$ may
well be rather different at the CMB and BBN epochs. In this case, Eq.
(\ref{neffxapprox}) at recombination reads
\begin{equation}
N_\nu^{\rm eff} = 3.0395 - (1-\beta) \left( \frac{z^{0}}{z^{\rm fin}}\right)^4
N_X^{BBN} + \Delta N_X^{CMB} \, \, ,
\label{neffxapproxcmb}
\end{equation}
where $N_X^{BBN}$ and  $\Delta N_X^{CMB}$ are the contribution of species
which are relativistic at the BBN and neutrino decoupling, and
recombination epochs, respectively.

We finally consider the case of massive active neutrinos. This nowadays
plausible scenario affects, in general, both the expectations for CMB
anisotropy and large scale structure formation. As it is seems more and
more clear from neutrino experiments on solar and atmospheric neutrinos, it
is unlikely that neutrino mass differences may be greater than $\sim 0.1$
eV \cite{pdg}, unless we enlarge the standard scenario introducing sterile
neutrino states. At the same time, it is also quite well established from
Tritium decay data that $\nu_e$ mass is bound to be smaller or, at most, of
the order of $1$ eV. It is therefore clear that in this scenario all
neutrino masses are completely negligible as far as their decoupling is
concerned. However if their values is as large as $\sim 1$ eV, they start
to be relevant as the temperature approaches the range relevant for CMB,
and the presence of a finite mass modify of course the neutrino
contribution to $N_\nu^{\rm eff}$. It is interesting to consider how the
effects of incomplete decoupling and QED thermal effects studied in the
previous Section would now affect the neutrino energy density. This is
conveniently parametrized by the (time dependent) quantity
\begin{equation}
\frac{\delta \rho_{\nu_\alpha}(m_\alpha)}{\rho_{\nu_\alpha}^0(m_\alpha)} \equiv
\frac{\rho_{\nu_\alpha}(m_\alpha) - \rho_{\nu_\alpha}^0(m_\alpha)}
{\rho_{\nu_\alpha}^0(m_\alpha)} \, \, ,
\label{varmass}
\end{equation}
which is defined as in Section 2, but for a neutrino with a finite mass
$m_\alpha$. Since for $m_\alpha \sim 1$ eV active neutrino are fully
relativistic for temperatures of the order of $MeV$, so we can still take
at the $e^{\pm}$ annihilation phase the results of Section 3, it is easy to see
that $\delta
\rho_{\nu_\alpha}(m)/ \rho_{\nu_\alpha}^0(m)$ is given by
\begin{equation}
\frac{\delta \rho_{\nu_\alpha}(m_\alpha)}{\rho_{\nu_\alpha}^0(m_\alpha)}=
10^{-4} \frac{\int_{0}^{\infty} dy~y^2~\sqrt{y^2 +
\left(\frac{m_{\alpha}}{m_e}\right)^2 x^2}\left(
 {\rm e}^y+1 \right)^{-1}
\sum_{i=0}^3 c^\alpha_i y^i}{\int_{0}^{\infty}
dy~y^2~\sqrt{y^2 + \left(\frac{m_{\alpha}}{m_e}\right)^2 x^2} \left(
 {\rm e}^y+1 \right)^{-1}} \, \, .
\label{deltarhomass}
\end{equation}

\begin{figure}{}
\psfig{file=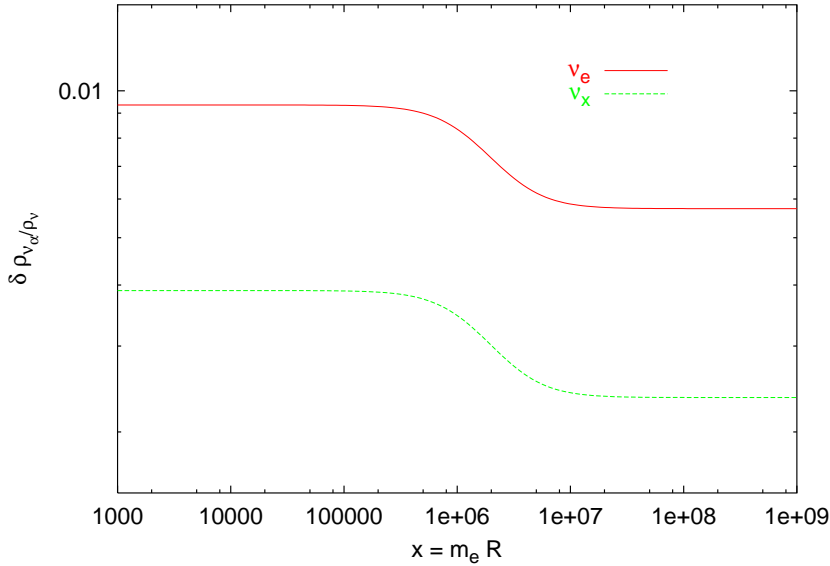,angle=-90,width=.8\textwidth}
\caption{Evolution of
$\delta \rho_{\nu_\alpha}(m_\alpha)/\rho_{\nu_\alpha}^0(m_\alpha)
\,$, for a neutrino mass $m_{\alpha} = 1\,$ eV (see text for further details).}
\label{fig1}
\end{figure}

In Figure 1, we plot $\delta \rho_{\nu_\alpha}(m_\alpha)/
\rho_{\nu_\alpha}^0(m_\alpha)$ for the reference choice $m_\alpha=1$
eV, using the coefficients $c^\alpha_i$ of Table 2. The $x$ range
corresponds to a variation of the temperature from $T \sim 1$ keV
till $T \sim 10^{-3}$ eV. We see that the effect of incomplete
decoupling and QED plasma masses for $e^{\pm}$ and $\gamma$ on
neutrino energy density decreases as the temperature becomes
comparable with the value of the neutrino mass. Notice that, from
(\ref{deltarhomass}), at very low values of $T$, when the neutrino
energy is dominated by the mass term, $\delta
\rho_{\nu_\alpha}(m_\alpha)/\rho^0_{\nu_\alpha}(m_\alpha)$ reaches an
asymptotic value, which is given by the change in the neutrino number
density due to the above effects, normalized with the number density
for a pure thermal Fermi-Dirac distribution.

\section{Conclusion}
\label{sec:concl}

In this paper we have considered how the two effects due to incomplete
neutrino decoupling and QED corrections to the $e^{\pm}$ and photon plasma
equation of state affect the effective neutrino degrees of freedom, a
crucial parameter for many cosmological observables.

The main result of our analysis, which have been carried out by numerically
solving the set of Boltzmann equations describing neutrino decoupling, is a
value for $N_\nu^{\rm eff}=3.0395$, for the standard case of three active
neutrino flavours. The issue is certainly not new, but we hope our study
will contribute to reach a better accuracy in the theoretical determination
of relativistic energy density of the Universe. In particular we have
stressed that there is quite an interesting interplay between the two
effects we have considered, so that at the level of accuracy of $0.005$ on
$N_\nu^{\rm eff}$, their corrections to the neutrino energy density cannot
be naively summed as they were fully independent.

We have also considered the less standard scenario of extra species
contributing to $N_\nu^{\rm eff}$ , and how they affect the calculation of
the thermal distortion of neutrino distribution, as well as the final value
for photon temperature after the $e^{\pm}$ annihilation phase.

Despite of the smallness of the corrections we have been concerned with in
this paper, nevertheless a careful analysis of data on the CMB anisotropy
spectrum at large multipoles may be able in the near future to reveal their
effects. Estimated sensitivity of Planck satellite experiment to
relativistic energy density is at least of the order of $\Delta N_\nu^{\rm
eff}\simeq 0.2$, but it is strongly improved, $\Delta N_\nu^{\rm eff}\simeq
0.005$ \cite{cmb2}, when including polarization measurements and strong
priors.

\section*{Acknowledgments}

S.P.~is supported by the European Commission by a {\it Marie Curie}
fellowship under the contract HPMFCT-2000-00445. M.P.~is supported by
the European Commission RTN programmes HPRN-CT-2000-00131, 00148, and
00152. In Munich, this work was partly supported by the Deut\-sche
For\-schungs\-ge\-mein\-schaft under grant No.\ SFB 375 and the ESF
network Neutrino Astrophysics.


\begin{thebibliography}{999}

\bibitem{kt}
E.W.~Kolb and M.S.~Turner, {\it The Early Universe}
(Addison-Wesley, 1990).
%%CITATION = NONE;%%

\bibitem{dicus}
D.A.~Dicus, E.W.~Kolb, A.M.~Gleeson,
E.C.~Sudarshan, V.L.~Teplitz and M.S.~Turner,
%``Primordial Nucleosynthesis Including Radiative, Coulomb, And Finite
%Temperature Corrections To Weak Rates,''
{\em Phys.\ Rev.\ D} {\bf 26} (1982) 2694.
%%CITATION = PHRVA,D26,2694;%%

\bibitem{heckler}
A.F.~Heckler,
%``Astrophysical Applications Of Quantum Corrections
%To The Equation Of State Of A Plasma,''
{\em Phys.\ Rev.\ D} {\bf 49} (1994) 611.
%%CITATION = PHRVA,D49,611;%%

\bibitem{Sh69}
V.F.~Shvartsman,
%``Density Of Relict Particles With Zero Rest Mass In The Universe,''
{\em Pisma Zh.\ Eksp.\ Teor.\ Fiz.\ } {\bf 9} (1969) 315
[{\em JETP Lett.\ } {\bf 9} (1969) 184].
%%CITATION = ZFPRA,9,315;%%

\bibitem{St77}
G.~Steigman, D.N.~Schramm and J.R.~Gunn,
%``Cosmological Limits To The Number Of Massive Leptons,''
{\em Phys.\ Lett.\ B} {\bf 66} (1977) 202.
%%CITATION = PHLTA,B66,202;%%

\bibitem{dhs}
A.D.~Dolgov, S.H.~Hansen and D.V.~Semikoz,
%``Non-equilibrium corrections to the spectra of massless neutrinos
%in the  early universe,''
{\em Nucl.\ Phys.\ B} {\bf 503} (1997) 426.
%[arXiv:hep-ph/9703315].
%%CITATION = HEP-PH 9703315;%%

\bibitem{dhs2}
A.D.~Dolgov, S.H.~Hansen and D.V.~Semikoz,
%``Nonequilibrium corrections to the spectra of massless neutrinos
%in the  early universe. (Addendum),''
{\em Nucl.\ Phys.\ B} {\bf 543} (1999) 269.
%[arXiv:hep-ph/9805467].
%%CITATION = HEP-PH 9805467;%%

\bibitem{emp}
S.~Esposito, G.~Miele, S.~Pastor, M.~Peloso and O.~Pisanti,
%``Non equilibrium spectra of degenerate relic neutrinos,''
{\em Nucl.\ Phys.\ B} {\bf 590} (2000) 539.
%[astro-ph/0005573].
%%CITATION = ASTRO-PH 0005573;%%

\bibitem{steigman}
G.~Steigman,
%``Precision neutrino counting,''
astro-ph/0108148.
%%CITATION = ASTRO-PH 0108148;%%

\bibitem{gnedin}
N.Y.~Gnedin and O.Y.~Gnedin,
%``Cosmological neutrino background revisited,''
{\em Astrophys.\ J.\ } {\bf 509} (1998) 11.
%arXiv:astro-ph/9712199.
%%CITATION = ASTRO-PH 9712199;%%

\bibitem{fdt}
B.~D.~Fields, S.~Dodelson and M.~S.~Turner,
%``Effect of neutrino heating on primordial nucleosynthesis,''
{\em Phys.\ Rev.\ D} {\bf 47} (1993) 4309.
%[arXiv:astro-ph/9210007].
%%CITATION = ASTRO-PH 9210007;%%

\bibitem{lotu}
R.E.~Lopez and M.S.~Turner,
%``An accurate calculation of the big-bang prediction for
%the abundance of  primordial helium,''
{\em Phys.\ Rev.\ D} {\bf 59} (1999) 103502.
%[arXiv:astro-ph/9807279].
%%CITATION = ASTRO-PH 9807279;%%

%\cite{Esposito:2000hh}
\bibitem{Esposito:2000hh}
S.~Esposito, G.~Mangano, G.~Miele and O.~Pisanti,
%``The standard and degenerate primordial nucleosynthesis versus recent  experimental data,''
{\em JHEP} {\bf 0009} (2000) 038.
%[arXiv:astro-ph/0005571].
%%CITATION = ASTRO-PH 0005571;%%

%\cite{izotov}
\bibitem{izotov}
Y.I.~Izotov and T.X.~Thuan, {\em Astrophys. J.} {\bf 500} (1998) 188.

%\cite{Hannestad:2001iy}
\bibitem{Hannestad:2001iy}
S.~Hannestad,
%``Oscillation effects on neutrino decoupling in the early universe,''
{\em Phys.\ Rev.\ D} {\bf 65} (2002) 083006.
%[arXiv:astro-ph/0111423]
%%CITATION = ASTRO-PH 0111423;%%

\bibitem{cmb}
R.~Bowen, S.H.~Hansen, A.~Melchiorri, J.~Silk, and R.~Trotta,
%The Impact of an Extra Background of Relativistic Particles on the
%Cosmological Parameters derived from Microwave Background Anisotropies
astro-ph/0110636.
%%CITATION = ASTRO-PH 0110636;%%

%\cite{Hannestad:2000hc}
\bibitem{Hannestad:2000hc}
S.~Hannestad,
%``New constraints on neutrino physics from Boomerang data,''
{\em Phys.\ Rev.\ Lett.\ } {\bf 85} (2000) 4203.
%[arXiv:astro-ph/0005018].
%%CITATION = ASTRO-PH 0005018;%%

%\cite{Orito:2000zb}
\bibitem{Orito:2000zb}
M.~Orito, T.~Kajino, G.J.~Mathews and R.N.~Boyd,
%``Neutrino degeneracy and decoupling: New limits from primordial
% nucleosynthesis and the cosmic microwave background,''
astro-ph/0005446.
%%CITATION = ASTRO-PH 0005446;%%

%\cite{Esposito:2000sv}
\bibitem{Esposito:2000sv}
S.~Esposito, G.~Mangano, A.~Melchiorri, G.~Miele and O.~Pisanti,
%``Testing Standard and Degenerate Big Bang Nucleosynthesis with BOOMERanG and MAXIMA-1,''
{\em Phys.\ Rev.\ D} {\bf 63} (2001) 043004.
%[arXiv:astro-ph/0007419].
%%CITATION = ASTRO-PH 0007419;%%

%\cite{Kneller:2001cd}
\bibitem{Kneller:2001cd}
J.P.~Kneller, R.J.~Scherrer, G.~Steigman and T.P.~Walker,
%``When Does CMB + BBN = New Physics?,''
{\em Phys.\ Rev.\ D} {\bf 64} (2001) 123506.
%[arXiv:astro-ph/0101386].
%%CITATION = ASTRO-PH 0101386;%%

%\cite{Hannestad:2001hn}
\bibitem{Hannestad:2001hn}
S.~Hannestad,
%``New CMBR data and the cosmic neutrino background,''
{\em Phys.\ Rev.\ D} {\bf 64} (2001) 083002.
%[arXiv:astro-ph/0105220].
%%CITATION = ASTRO-PH 0105220;%%

\bibitem{cmb2}
R.E.~Lopez, S.~Dodelson, A.~Heckler and M.S.~Turner,
%``Precision detection of the cosmic neutrino background,''
{\em Phys.\ Rev.\ Lett.\ }  {\bf 82} (1999) 3952.
%[arXiv:astro-ph/9803095].
%%CITATION = ASTRO-PH 9803095;%%

\bibitem{SZ}
U.~Seljak and M.~Zaldarriaga,
%``A Line of Sight Approach to Cosmic Microwave Background Anisotropies,''
{\em Astrophys.\ J.\ }  {\bf 469} (1996) 437.
%[arXiv:astro-ph/9603033].
%%CITATION = ASTRO-PH 9603033;%%

\bibitem{herrera}
M.A.~Herrera and S.~Hacyan,
{\em Astrophys.\ J.\ } {\bf 336} (1989) 539.
%%CITATION = NONE;%%

\bibitem{rana}
N.C.~Rana and B.~Mitra,
%``Effect Of Neutrino Heating In The Early Universe On
%Neutrino Decoupling Temperatures And Nucleosynthesis,''
{\em Phys.\ Rev.\ D} {\bf 44} (1991) 393.
%%CITATION = PHRVA,D44,393;%%

\bibitem{dt}
S.~Dodelson and M.S.~Turner,
%``Nonequilibrium neutrino statistical mechanics in the expanding universe,''
{\em Phys.\ Rev.\ D} {\bf 46} (1992) 3372.
%%CITATION = PHRVA,D46,3372;%%

\bibitem{df}
A.D.~Dolgov and M.~Fukugita,
%``Nonequilibrium Effect Of The Neutrino Distribution
%On Primordial Helium Synthesis,''
{\em Phys.\ Rev.\ D} {\bf 46} (1992) 5378.
%%CITATION = PHRVA,D46,5378;%%

\bibitem{madsen}
S.~Hannestad and J.~Madsen,
%``Neutrino decoupling in the early universe,''
{\em Phys.\ Rev.\ D} {\bf 52} (1995) 1764.
%[arXiv:astro-ph/9506015].
%%CITATION = ASTRO-PH 9506015;%%

\bibitem{nicolao}
N.~Fornengo, C.W.~Kim and J.~Song,
%``Finite temperature effects on the neutrino decoupling in
%the early  universe,''
{\em Phys.\ Rev.\ D} {\bf 56} (1997) 5123.
%[arXiv:hep-ph/9702324].
%%CITATION = HEP-PH 9702324;%%

\bibitem{bbn}
S.H.~Hansen, G.~Mangano, A.~Melchiorri, G.~Miele, and O.~Pisanti,
%``Constraining neutrino physics with BBN and CMBR.''
{\em Phys.\ Rev.\ D} {\bf 65} (2002) 023511.
%[arXiv:astro-ph/0105385].
%%CITATION = ASTRO-PH 0105385;%%

\bibitem{cmb3}
A.R.~Zentner and T.P.~Walker,
%``Constraints on the Cosmological Relativistic Energy Density''
{\em Phys.\ Rev.\ D} {\bf 65} (2002) 063506.
%[arXiv:astro-ph/0110533].
%%CITATION = ASTRO-PH 0110533;%%

\bibitem{pdg}
D.E.~Groom {\it et al.}  [Particle Data Group Collaboration],
%``Review Of Particle Physics,''
{\em Eur.\ Phys.\ J.\ C} {\bf 15} (2000) 1.
%%CITATION = EPHJA,C15,1;%%


\end{thebibliography}
\end{document}